\documentclass[aps,prl,twocolumn,superscriptaddress]{revtex4-1}

\usepackage[colorinlistoftodos]{todonotes}
\usepackage{epsfig}% Include figure files
\usepackage{amsmath}
\usepackage{amssymb}
\usepackage{bm}% bold math
\usepackage{siunitx}

\newcommand{\be}{\begin{equation}}
\newcommand{\ee}{\end{equation}}
\newcommand{\bea}{\begin{eqnarray}}
\newcommand{\eea}{\end{eqnarray}}
\newcommand{\om}{\omega}

\begin{document}

%\preprint{}

\title{Recoil-sensitive lithium interferometer without a subrecoil sample}



\author{Kayleigh Cassella}
\email{kcassell@berkeley.edu}
\author{Eric Copenhaver}
\author{Brian Estey}
\affiliation{Department of Physics, 366 LeConte Hall, Berkeley, CA 94720, USA} 
\author{Yanying Feng}
\affiliation{Joint Institute for Measurement Science, State Key Laboratory of Precision Measurement Technology \& Instruments, Department of Precision Instruments, Tsinghua University, Beijing, China, 100084}
\author{Chen Lai}
\affiliation{Department of Mathematics,
University of California, San Diego, La Jolla, CA 92093-0404,
USA}
\author{Holger M\"uller}
\altaffiliation{Chemical Sciences Division, Lawrence Berkeley National Lab}
\affiliation{Department of Physics, 366 LeConte Hall, Berkeley, CA 94720, USA} 

\date{\today}

\begin{abstract}

We report simultaneous conjugate Ramsey-Bord\'e interferometers with a sample of low-mass (lithium-7) atoms at 50 times the recoil temperature. We optically pump the atoms to a magnetically insensitive state using the $2S_{1/2} - 2P_{1/2}$ line. Fast stimulated Raman beam splitters address a broad velocity class and unavoidably drive two conjugate interferometers that overlap spatially. We show that detecting the summed interference signals of both interferometers, using state labeling, allows recoil measurements and suppression of phase noise from vibrations. The use of ``warm" atoms allows for simple, efficient, and high-flux atom sources and broadens the applicability of recoil-sensitive interferometry to particles that remain difficult to trap and cool. 
% * <kcassell@berkeley.edu> 2016-08-14T20:56:52.009Z: ... with a light atom () and at the same time
%
% ^.
\end{abstract}

% \pacs{37.10.De}

\maketitle
% 
%\section{Introduction}

In a light-pulse atom interferometer, laser pulses with wavenumber $k$ direct matter waves along a superposition of trajectories and recombine them to reveal the phase difference between paths \cite{CroninReview}. They are used for inertial sensing \cite{Peters,Geiger}, gravity gradiometry \cite{McGuirk} and tests of fundamental physics \cite{LVGrav,Hohensee2012,EEP1,EEP2,EEP3,EEP4,EEP5,EEP6,EEP7,Chameleons,TinoG}. Ramsey-Bord\'e interferometers, in particular, measure the mass $m$ of an atom through the kinetic energy $\hbar \om_r=\hbar^2k^2/(2m)$ it gains after recoiling from the interaction with a photon ($\hbar$ is the reduced Planck constant). They can help redefine the kilogram \cite{CCC,Bouchendira2013} and determine the fine-structure constant \cite{Weiss1994,Wicht,Bouchedira2011,Jamison,Estey2015}, thereby testing the Standard Model \cite{Gabrielse,PenningTrapSME}. The recoil frequency $\om_r$, and therefore the signal, scales inversely with mass. Light atomic species have been used in supersonic atomic-beam interferometers \cite{Miffre,Jac}, but remain difficult to cool below the recoil temperature $T_r$ where the average thermal speed equals the recoil velocity. This makes it impossible to spatially resolve the interferometer outputs, which is required for direct rejection of common-mode inertial signals with phase extraction methods \cite{SCI,Noisered,Bayanal}. 

\begin{figure}[h!]
\centering
\epsfig{file=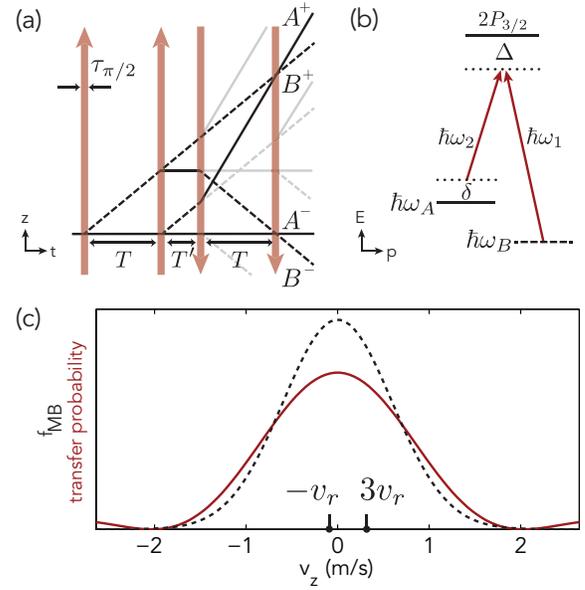,width=0.42\textwidth}
\caption{\label{trajectories} Ramsey-Bord\'e interferometry with high temperatures: (a) Space-time trajectories of atoms in Ramsey-Bord\'e interferometers, neglecting gravity. Solid and dashed lines indicate internal states of the atom, for example hyperfine ground states. Interfering trajectories are shown in black and non-interfering outputs are shown in light gray. Arrows on the light pulses represent the effective wave vector. (b) Energy levels and frequencies involved in Raman transitions. (c) The bandwidth of the atomic response to a $\pi/2$ pulse (solid red) is inversely proportional to pulse duration, while the velocity width of the Maxwell-Boltzmann distribution along the Raman axis (dashed black) is proportional to the square root of the temperature (here 300 $\mu$K). The 160-ns pulses cover a large velocity class, including the speeds that Doppler shift the third and fourth pulses onto resonance for each conjugate interferometer.}
\end{figure}

Here, we demonstrate recoil-sensitive interferometry with a sample of lithium-7 atoms well above the atomic recoil temperature ($50T_r$), the first interferometer with laser-cooled lithium atoms or any atom lighter than sodium-23 \cite{KasevichChu}. Fast Raman transitions \cite{BiedermannWarm} ($\tau_{\pi/2}\!=$160 ns) address the ensemble's large Doppler spread and simultaneously drive overlapped conjugate Ramsey-Bord\'e interferometers. Superimposing simultaneous conjugate interferometers suppresses effects from two-photon detuning and unwanted inertial signals, such as vibrations. Our measurement sensitivity benefits from lithium's high recoil frequency of $\omega_r\!=\!2\pi\times$63\,kHz (compared to $2\pi\times$2\,kHz for cesium) and the absence of time-consuming additional cooling \cite{AllOptical} or lossy velocity selection  \cite{vsel} steps that reduce sample size and precision. The lithium isotopes present an attractive pair for testing Einstein's Equivalence Principle using light-pulse atom interferometry \cite{nuclear}. This work broadens the applicability of recoil-sensitive interferometry to other particles; electrons \cite{eCooling}, for example, boast GHz-recoil frequencies and would enable observation of relativistic effects \cite{eLorentz,CCC}. 

%While inertially-sensitive light-pulse Mach-Zehnder interferometers have recently been implemented with room-temperature atoms \cite{BiedermannWarm}, interferometers measuring the atomic recoil have relied on laser-cooled samples below the recoil temperature, $T_r=\hbar^2k^2/k_Bm$ (where $\hbar$ is the reduced Planck constant, $k$ is the radiation wavenumber, $m$ is the atomic mass, and $k_B$ is the Boltzmann constant), in which the velocity spreads are smaller than the recoil velocity, $v_r =\hbar k/m$. %Spatially-resolved detection of simultaneous conjugate interferometer outputs allows for correlation of the two phases and for rejection of unwanted, common-mode inertial signals \cite{SCI}.
% However, while subrecoil cooling is possible for heavy atoms like rubidium and cesium, their low recoil frequencies necessitate large-momentum-transfer beam splitters that complicate systematic effects \cite{Multiport}. Such subrecoil samples also require velocity selection or additional cooling steps , both of which reduce measurement sensitivity through lower atom numbers and increased cycle times. Performing recoil-sensitive interferometry with warm samples of light particles, which offer inherently high recoil frequencies, obviates the need for these techniques.

%\section{Conjugate interferometers}

Figure \ref{trajectories}(a) shows the trajectories of an atom in a Ramsey-Bord\'e sequence. Atom-light interactions are used to split, redirect and interfere the atomic matter waves. The Ramsey-Bord\'e sequence consists of four $\pi/2$ (beam-splitter) pulses, so that the lowest interferometer arm remains stationary. %, but results in a 50\% loss of input atoms at both the second and third pulses and an overall reduction in maximum contrast between interferometer outputs $A^{\,-}$ and $B^{\,-}$. However, 
The outputs of the second pulse that do not contribute to $A^{\,-}$ and $B^{\,-}$ may form another conjugate (upper) interferometer with final outputs $A^+$ and $B^+$.
%For an atom beginning in state A^{(\pm)} prior to the interferometry pulse sequence, the probability of emerging  in  state B^{(\pm)} oscillates as $P_{B^{(\pm)}} = D(1-C_{(\pm)}\cos{\Delta\phi^{(\pm)}})$ where $D$ is an overall offset, C_{(\pm)} is the fringe contrast and $\Delta\phi^{(\pm)}$ is the phase difference between the arms in an interferometer. Using standard methods \cite{HoganLPAI}, $\Delta\phi^-$ ($\Delta\phi^+$) for the lower (upper) interferometer is calculated to second order in $T$ as
In each interferometer, the probability of detecting the atoms at one output depends on the phase difference between the arms of the interferometer, which we denote $\Delta \phi^-$ ($\Delta \phi^+$) for the lower (upper) interferometer. Using standard methods \cite{HoganLPAI}, $\Delta\phi^\pm$ is calculated to second order in $T$ as 
\begin{equation}\label{RBIphase}
\Delta \phi^\pm =\pm 8 \omega_r T - 2k a_z T(T+T')-2\delta T% = \pm\phi_{r}+\phi_c\,.
\end{equation}
The first term arises from the atomic kinetic energy, the second from any acceleration $a_z$ (such as gravity and vibrations) along the laser beam axis, where the average wave number of the counter-propagating beams is $k=(k_1+k_2)/2$, and the third from the detuning of the laser frequencies from two-photon resonance in the absence of AC stark shifts, $\delta=\omega_1-\omega_2-(\omega_A-\omega_B)$ \footnote{We call this term a two-photon detuning loosely. It is the difference between the Raman laser frequency difference $\omega_1-\omega_2$ and the hyperfine splitting $\omega_A-\omega_B$ during free evolution (see Fig. \ref{trajectories}(b)). That is, our so-called two-photon detuning does not include AC Stark shifts of the internal hyperfine energies induced when the Raman light dresses the atoms, but it does include other shifts that persist during free evolution such as Zeeman shifts.}. %The phase terms have been regrouped such that $\phi_r$ represents the phase proportional to the recoil frequency whereas $\phi_c$ represents the phase contribution common to both interferometers. 

%\subsection{Overlap at high atom temperature}  

%If all overlapping outputs $A^{\,-}$\!, $A^+$\!, $B^{\,-}$\!, and $B^+$ are in the same internal state, interference is undetectable. Without loss of contrast and without sacrificing atom counting statistics through further cooling or velocity selection, we use two-photon Raman transitions as beam splitters to couple the atoms' external and internal degrees of freedom and achieve spectral resolution of the outputs. 

The interferometers in Fig. \ref{trajectories}(a) share the first and second beam-splitter pulses. For the third and fourth pulse, the lower interferometer requires e a transition coupling $|F=2,p=0\rangle\rightarrow|F=1,p=-2\hbar k\rangle$, and the upper interferometer requires coupling  $|F=1,p=+2\hbar k\rangle\rightarrow|F=2,p=+4\hbar k\rangle$. Reversing the effective wave vector of the beam splitters for the second pulse pair accomplishes both of these couplings. In principle, they are distinguished by a Doppler shift of $8\omega_r$ due to the speed difference between the lower and upper interferometer, as marked in Fig. \ref{trajectories}(c). Low-bandwidth beam-splitter pulses for atom interferometers typically resolve this frequency difference, but % and address only one of the transitions at the third and fourth pulses, requiring an additional frequency to drive both transitions and close both interferometers. 
the high-bandwidth pulses we use to address a broad velocity class simultaneously address both transitions, unavoidably closing both interferometers. %One interferometer's outputs may be detected separately if they are state-labeled as with Raman transitions, but 
The two interferometer's outputs ports ({\it e.g.} $B^{\,-}$\!, $B^+$) overlap spatially since the samples thermally expand faster than the interferometers separate.
 
% One interferometer's outputs may be detected separately if they are state-labeled, as with Raman transitions, though outputs of conjugate interferometers ($B^{\,-}$\!, $B^+$) remain overlapped.
%\subsection{Fringes of overlapped conjugate interferometers}

We recover the recoil signal by using Raman beam splitters, which allow us to use  state-dependent detection of the sum of signals from the lower and upper interferometers. Beginning in the $|F=2\rangle$ ground state (state $A$) prior to the interferometry pulse sequence, the probability for an atom to emerge from the interferometer in the $|F=1\rangle$ ground state (state $B$) oscillates as:
\be\label{eq:4}
P_B = D\big[ 1 - C_-\cos(\Delta \phi^-) - C_+\cos(\Delta \phi^+)\big], %\nonumber \\
% & = & A - C_-\sin[2\pi(8\omega_r T - 2\delta T )] \nonumber \\ && + C_+\sin[2\pi(8\omega_r T+ 2\delta T )]
\ee
%where $C\!_{_\pm}$ are the fringe contrasts of each interferometer and $D$ is an overall offset. For equal contrasts, the signal simplifies to $\cos(\phi_r)$ modulated by $\cos(\phi_c)$. 
where $C_\pm$ are the fringe contrasts of each interferometer and $D$ is an overall offset. For approximately equal contrasts, $C_+=C_-\equiv C/2$, the signal simplifies to:
\be
P_B= D [1-C\cos(2k a_z T(T+T') + 2\delta T)\cos(8\omega_r T)].
\label{eq:6}
\ee

\begin{figure}[t]
\centering
\epsfig{file=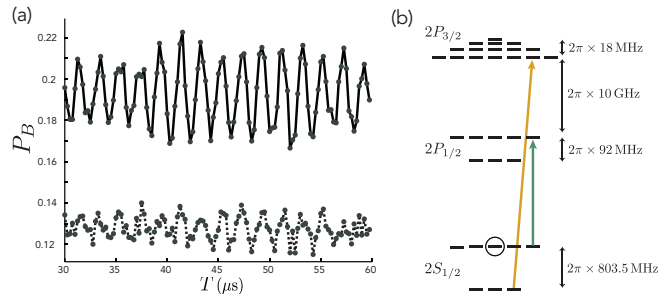,width=0.48\textwidth}
\caption{\label{OP} Optical pumping: (a) Interference fringes without optical pumping (lower dashed curve) and with optical pumping (upper solid curve). Each gray point on the traces is the average of 5 experimental shots and error bars are omitted for clarity. (b) Optical pumping on lithium's $D_1$ line with $\pi$ light (green arrow) results in a dark state at $|F\!=\!2,m_F\!=\!0\rangle$ (black circle). Atoms that decay to $|F=1\rangle$ are recovered by 3D MOT repump light (yellow arrow). Each dash represents a unique magnetic sublevel.}
\end{figure}

%\section{Setup}

%\subsection{Trapping and cooling}

Our setup is similar to the one previously described in Ref. \cite{LIcooling} but without the polarization gradient lattice used for sub-Doppler cooling. We heat lithium to 400$^\circ$C and trap the vapor in a two-dimensional (2D) magneto-optical trap (MOT). A push beam tuned near resonance sends the atoms through a differential pumping tube into the interferometry chamber, where approximately 15 million atoms are trapped in a three-dimensional (3D) MOT. After lowering the intensities of both the cooling and repumping light and moving the detuning closer to resonance, the cloud reaches a final temperature of roughly 300\,$\mu$K.

To define a quantization axis for optical pumping and Raman transitions, we %use three axes of bias magnetic field coils to zero the field within 10 mG along $\hat{x}$ and $\hat{y}$ and to 
apply a 1-G  bias magnetic field along the $\hat{z}$ axis. Despite the 250-$\mu$s decay of the current in the anti-Helmholtz MOT coils, the quadrupole field remains appreciable for milliseconds due to eddy currents in the steel vacuum chamber. We use the 3D MOT beams as optical molasses to limit the thermal expansion of the cloud while the eddy currents decay. No polarization gradient cooling occurs during this step due to the small detuning of the 3D MOT beams from the unresolved $D_2$ line ($2P_{3/2}$ state) \cite{LIcooling}.

%\subsection{Optical pumping}

After the optical molasses, the atoms are distributed among the five non-degenerate magnetic sublevels of the $|F=2\rangle$ ground-state manifold. This leads to magnetic dephasing since the Ramsey-Bord\'e interferometer phase depends on the internal energies through the $\delta$ term. Interferometer experiments often select atoms in the desired magnetic sublevel by transferring them to the other hyperfine state with a microwave and blowing away the remaining populations with resonant light. The unresolved $D_2$ line in lithium, however, precludes the efficient cycling transitions required to impart the large momentum needed for such blow-away beams. Furthermore, this selection process is lossy, as large atomic populations are sacrificed to the blow-away beam. 

To avoid the magnetic dephasing from atoms in different magnetic sublevels, we optically pump the sample to the magnetically insensitive $|F=2,m_F=0\rangle$ state by taking advantage of the selection rule that prohibits $m_F=m_F'=0$ transitions when $\Delta F=0$. Once the magnetic field gradient decays below 1\,G/cm (after 1.5\,ms of optical molasses), we send 3\,mW of light tuned within a linewidth ($\Gamma/2\pi=$ 5.87 MHz) of the $|F=2\rangle$ to $|F^\prime=2\rangle$ transition on the well-resolved $D_1$ line ($2P_{1/2}$ state). The optical pumping light is $\pi$ polarized along $\hat{z}$ and has a 3.6-mm Gaussian waist. Unlike the $D_2$ line, lithium's $D_1$ line has a resolved hyperfine structure (see Fig. \ref{OP}(b)). Optical pumping on the $D_1$ line therefore avoids the slightly off-resonant transitions ubiquitous on the $D_2$ line \cite{LiOP}. In each of the six 3D MOT beams, we use 1.5 mW of $D_2$ MOT repump light to recover atoms that decay to $|F=1\rangle$. We tune the repump frequency closer to resonance, optimizing for optical pumping efficiency. After 50\,$\mu$s of optical pumping, more than $80\%$ of the atoms occupy the dark state.

%\subsection{Raman pulses}

%\subsection{Imaging}

Figure \ref{OP} displays the efficacy of the optical pumping for interferometry. Without optical pumping, the recoil fringes have low contrast, a low signal-to-noise ratio, and decohere more rapidly, limiting the maximum interrogation time and sensitivity. Preparation to the magnetically insensitive $|F=2,m_F=0\rangle$ state before interferometry increases the contrast and signal-to-noise ratio by more than a factor of 2 at short interrogation times. Optical pumping also makes the fringes visible at longer interrogation times. 

After optical pumping, we measure the fringes by varying the separation time $T$ while keeping $T^\prime = 10\,\mu $s and $\delta$ constant but small compared to $\omega_r$. To close the interferometers, we reverse the direction of the Raman beams for the second pulse pair using an electro-optic modulator (see Supplemental Material). For normalized detection, we use a new imaging technique that captures two images during a single exposure (see Supplemental Material). 

Figure \ref{fringes} shows the summed interference fringes obtained from the simultaneous conjugate Ramsey Bord\'e interferometers. As seen in Eq. \ref{eq:6}, they can be described by a fast oscillation at a frequency of $8\om_r$ within an envelope function that oscillates slowly at a frequency set by the two-photon detuning $2\delta$, in addition to accelerations of the atoms $a_z$. Here, the two-photon detuning term dominates over phases induced by acceleration, because we operate our interferometer perpendicular to gravity and at short interrogation times. Fig. \ref{fringes}(b) shows the fast component of the summed fringes. We fit the fringes using a least-squares method to the functional form in Fig. \ref{fringes}(b). The confidence interval in the fit constitutes a 32 ppm recoil measurement in 2 hours. After averaging across 10 such data sets with varying $\delta$, we reached a precision of 10 ppm. The phase sensitivity of the fit corresponds to a sensitivity roughly 50 times larger than the shot-noise limit. %While data acquisition schemes that concentrate only on fringes at high phase would improve the sensitivity and averaging, we focus on the full temporal signal to emphasize the techniques we develop.

The noise observed in the data is due mostly to laser noise, as we have confirmed by numerical simulations adapted from previous studies of noise in Ramsey-Bord\'e interferometers \cite{Multiport}. The linewidth of the Raman laser ($\gamma/2\pi\approx1$ MHz) is sizable compared to the small magnitude of the single-photon detuning ($\Delta/2\pi=210$ MHz) and creates pulse-to-pulse fluctuations of the two-photon Rabi frequency, which result in noise significantly larger than the shot-noise-limited sensitivity. %Fluctuations in the laser frequency modify the pulse transfer probability, which contributes to phase uncertainty in the fit. 

\begin{figure}[t]
\centering
\epsfig{file=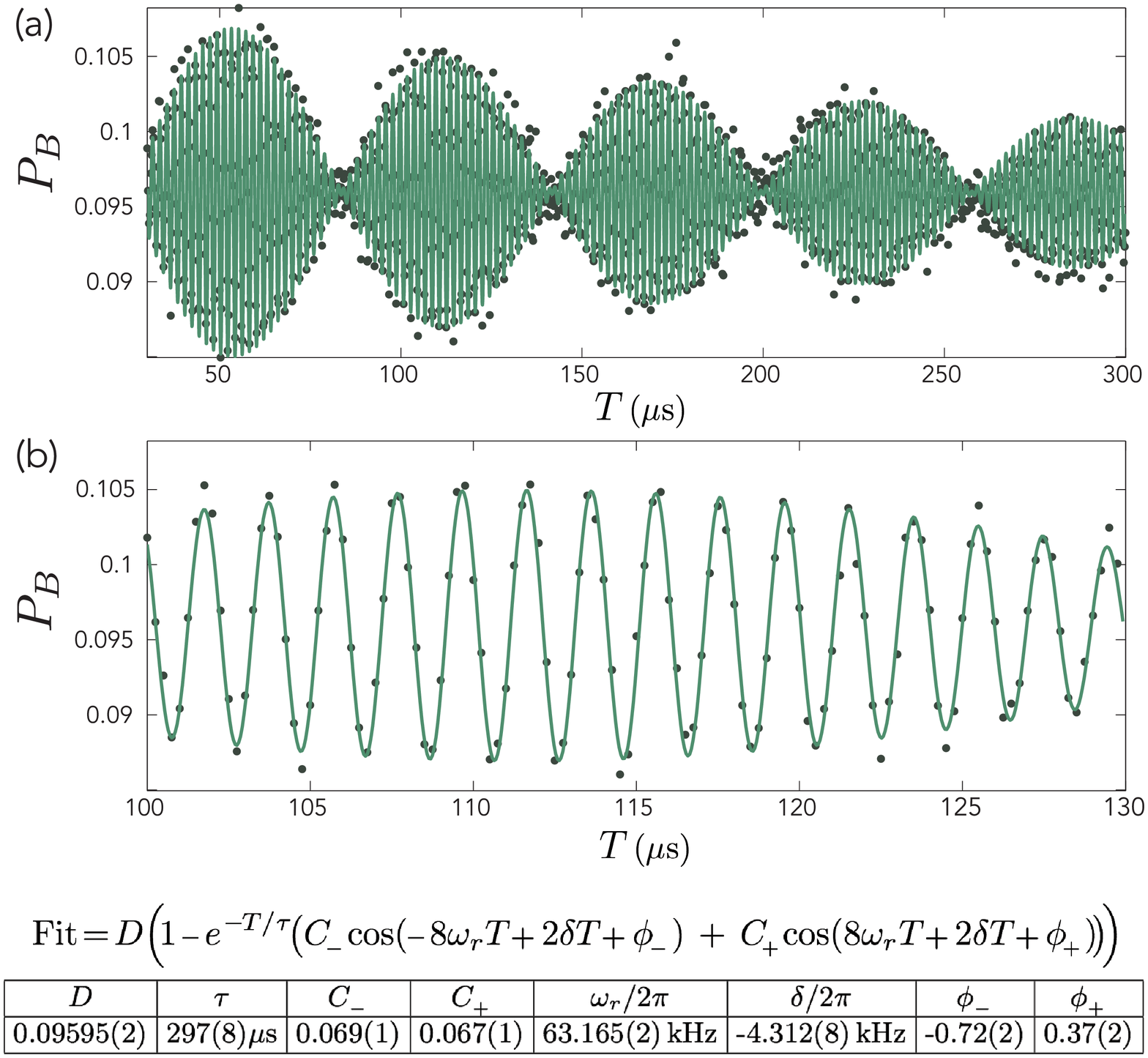,width=0.48\textwidth}
\caption{\label{fringes} Beating interference of overlapped interferometers: (a) The probability of detecting atoms in the $|F=1\rangle$ state oscillates, beating due to a non-zero $\delta\!=\!-2\pi\!\times\!4.3$ kHz. Each point is the average of 5 experimental shots with error bars omitted for clarity. Fitting (in green) yields $\omega_r = 2\pi\times (63.165\pm 0.002$ kHz). (b) Closer inspection reveals the fast recoil component of the fringes. The table below shows results of the fit with 1-$\sigma$ precision.}
\end{figure}

%Our coherence time is currently limited by magnetic dephasing of the $m_F=0$ atoms that results from inhomogeneous quadratic Zeeman shifts which survive after the optical molasses. 
The coherence time of the interferometer is not yet limited by thermal expansion out of the Raman beam but instead by magnetic dephasing of the $m_F=0$ atoms. The magnetic field gradient that survives after the optical molasses gives rise to inhomogeneous quadratic Zeeman shifts, leading to an interferometer phase dependent on an atom's position in the cloud. We are able to reduce the magnetic gradient by extending the optical molasses time to 5 ms and, with half the remaining gradient, the interference contrast indeed decays at half the rate. Magnetic gradient compensation would lead to longer coherence times and improved sensitivity. At a conservatively projected $T$=1\,ms, we estimate the shot-noise-limited sensitivity with $10^7$ atoms to be 100 ppb/$\sqrt{\rm{Hz}}$. Implementing sub-Doppler cooling techniques \cite{LIcooling,GrayMolasses} to reach a temperature of 40 $\mu$K (approximately $8 T_r$) would improve the sensitivity by $\sqrt{50/8}\sim 3$, but still require the techniques in this paper. 

%\section{Summary and Outlook} 

Phase shifts due to vibrations cancel when the fringes are summed in our detection scheme, as they enter the conjugate interferometers with opposite sign. The only effect of vibrations is then an amplitude modulation of the fringes. Consider Eq. (\ref{RBIphase}) with a stochastic, Gaussian-distributed $a_z$ with 0 mean and standard deviation $\sigma$. When $2k \sigma T(T+T') \ll \pi$, the effect of such vibrations is a modulation of the interference contrast, which decreases proportionally to $a_z^2$. Other interferometers operating on a similar optical table without vibration isolation accrue phase shifts much less than $\pi$ due to vibrations, even at $T=10$ ms \cite{Paul}. Lithium's high recoil frequency allows us to take sensitive data at $T<10\,$ms, and therefore to make full use of the common-mode rejection of vibration-induced signals. %\todo{should we not make the following claim without more evidence, or should we include supplemental material on it?} Should vibrations dominate with $2k \sigma T(T+T') \gg \pi$, the fringe amplitude is modulated with random magnitude and sign. Even still, points where $\cos(8\omega_rT)=0$ are still respected and $|\cos(8\omega_rT)|$ fringes develop in the noise of the detected atom number.

This demonstration of interferometry opens the door to recoil measurements with other particles that are difficult to cool to subrecoil temperatures, such as electrons. Electrons, whose recoil frequency is on the order of GHz, are susceptible to relativistic effects and consequently a recoil-sensitive measurement can be used to measure Lorentz contraction \cite{eLorentz}. % or search for signals of {\it CPT}- and Lorentz-violation \cite{PenningTrapSME}. 
While Kapitza-Dirac scattering has been proposed to realize matter-wave beam splitters for electrons in a Ramsey-Bord\'e interferometer \cite{eRBI}, any vibrations or nonzero two-photon detuning will modify the phase $\Delta\phi^-$ for a single Ramsey-Bord\'e. As we have shown in this work, the inclusion of the simultaneous conjugate interferometer ($\Delta\phi^+$) recovers the recoil phase independently of a two-photon detuning even when the outputs of conjugate interferometers are spatially unresolved, as would the case for electron plasmas in a Penning-Malmberg trap \cite{eCooling}. The required spectral resolution for detection could be achieved with bichromatic Kapitza-Dirac pulses. Bichromatic pulses with very large intensity have been proposed to impart momentum to an electron while inducing a spin flip \cite{McGregorBichromatic} and hence couple the electron's external and internal degrees of freedom. With such beam splitters acting on a spin-polarized sample and spin-dependent detection, the techniques we demonstrate in this work pave the way for a recoil-sensitive electron interferometer.

In summary, we demonstrate recoil-sensitive Ramsey-Bord\'e interferometry with laser-cooled lithium-7 at 300 $\mu\text{K}$ ($50T_r$). The large Doppler spread of the sample is addressed with fast pulses, driving simultaneous conjugate interferometers with nearly equal contrast. Even with non-zero two-photon detuning, the interference fringes allow for the determination of the recoil frequency independent of two-photon detuning and vibrations. We suppress first-order magnetic dephasing and extend the coherence time by optically pumping the atoms to the magnetically insensitive $|F=2,m_F=0\rangle$ state using lithium-7's well-resolved $D_1$ line. Our results relax cooling requirements for recoil interferometry, allowing for increased precision through high experimental repetition rates \cite{BiedermannWarm,Biedermann14}. Extending these techniques %, as described here for the case of electrons, 
would allow for recoil-sensitive interferometry with atoms and other particles that have thus far been excluded from such experiments.

\acknowledgments We thank Paul Hamilton, Philipp Haslinger, Matt Jaffe, Geena Kim, Richard Parker, Simon Budker, Quinn Simmons, Dennis Schlippert, and Daniel Tiarks for help with the experiment and discussions. This work relies on a diverse and inclusive environment for all contributing scientists. This material is based upon work supported by the National Science Foundation under CAREER Grant No. PHY-1056620, the David and Lucile Packard Foundation, and National Aeronautics and Space Administration Grants No. NNH13ZTT002N, No. NNH10ZDA001N-PIDDP, and No. NNH11ZTT001.

%% Create the reference section using BibTeX:

\newpage

\section{Supplemental Material}
\begin{figure}[b]
\centering
\epsfig{file=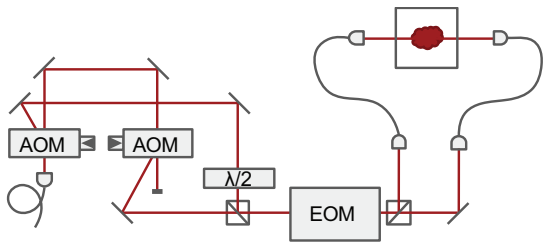,width=0.4\textwidth}
\caption{\label{Raman} Raman beam setup: 400-MHz AOM's shift the frequencies of the Raman beams to match the hyperfine splitting. The EOM reverses the propagation direction of the Raman beams with frequencies $\omega_1$ and $\omega_2$ between the second and third pulses, during $T'$.}
\end{figure}

% \subsection{Raman beams}
{\bf Raman beams:} 
%The Raman beams are generated from a diodeThe single-photon detuning of the Raman pair from the $D_2$ line is $\Delta=2\pi\times 210$ MHz. 
Light enters the setup as shown in Fig. \ref{Raman}, red-detuned with a single-photon detuning of $\Delta=2\pi\times 210$ MHz from the crossover peak of the two hyperfine transitions on the $D_2$ line. Two acousto-optic modulators (AOM's) operating near 400 MHz (IntraAction ATM-4001A1) generate a frequency difference close to the 800-MHz ground state hyperfine splitting, each shifting either up or down by $(\omega_A-\omega_B+\delta)/2$. 
Approximately 30\,mW of $\omega_1$ and 15\,mW of $\omega_2$ coincide at the 2-mm diameter cloud in beams of 2.1-mm Gaussian waist. The beams realize a lin$\perp$lin geometry, with one polarized along $\hat{x}$ and the other along $\hat{y}$. The relatively small single-photon detuning allows for a high two-photon Rabi frequency $\Omega_R\sim 2\pi\times 1.6$ MHz and a short pulse duration. We drive a $\pi$ pulse in 320\,ns with $\sim 30\%$ efficiency ($\tau_{\pi/2}$=160 ns). These large-bandwidth pulses address a considerable fraction of the atoms, whose two-photon resonance conditions are Doppler-broadened from the thermal velocity spread.
%The frequency difference of the beams becomes $\omega_A-\omega_B+\delta$. 

For Ramsey-Bord\'e interferometry based on Raman transitions, we must switch the propagation directions of $\omega_1$ and $\omega_2$ between the second and third pulses in order to close the interferometer. To achieve this, we orthogonally polarize $\omega_1$ and $\omega_2$ and overlap them before passing them through an electro-optic modulator (EOM) that acts as a voltage-controlled wave plate. A polarizing beam splitter following the EOM separates the frequencies and directs the light to fibers that send the beams to the atoms from opposing directions. By switching the EOM voltage from 0 V to 215 V during $T'$, we rotate the polarizations of the frequencies by $90^\circ$ and consequently reverse the Raman wave vectors.

\begin{figure}[b]
\centering
\epsfig{file=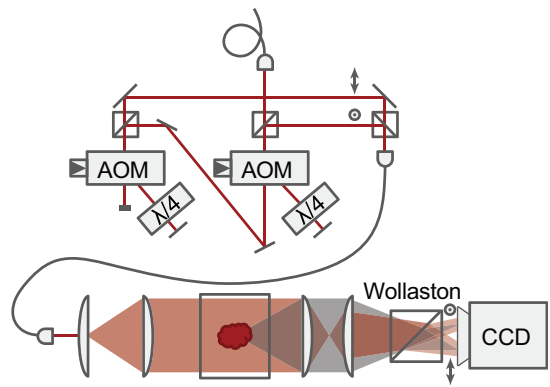,width=0.4\textwidth}
\caption{\label{imaging} Wollaston-based normalized imaging setup: orthogonal polarizations of the same imaging frequency are controlled by separate AOM's. The orthogonal polarizations form distinct absorption images during a single exposure of a CCD camera. The imaging beam is shown in red, while gray indicates the shadow that comprises the absorption imaging signal.}
\end{figure}

% % --- DON'T DELETE THIS PHANTOM FIGURE --- %
% \begin{figure}[b]
% \centering
% \epsfig{file=WollastonImaging.eps}
% \caption{\label{SuchMystery} }
% \end{figure}
% % --- THE WOLLASTON FIG ABOVE WON'T SHOW WITHOUT IT ... ? --- %

%\subsection{Normalized detection with single exposure}
{\bf Normalized detection with a single exposure:} Shot-to-shot fluctuations in the total atomic population lead to noise in the interference signal. This effect can be lessened by normalizing each shot of the experiment to the number of atoms trapped on each shot. Interferometers based on cesium or rubidium can rely on state-selective cycling transitions to push a single hyperfine state aside and image spatially resolved populations. Lithium's unresolved hyperfine structure precludes efficient cycling transitions, so another solution is needed. The long readout time of most CCD cameras makes it impossible to take successive exposures of hot samples, since the sample will have diluted by the beginning of the second exposure due to thermal expansion.

We normalize our detection with two state-selective images during a single 190-$\mu$s exposure of a CCD camera (PCO pixelfly) as described in Fig. \ref{imaging}. Light locked to the $D_2$ crossover passes twice through one of two 200-MHz AOM's (Crystal Tech AOMO 3200-125), each of which produces an orthogonal polarization. Both frequencies address the $|F=1\rangle$ state on the $D_2$ line. During the first 90 $\mu$s of the exposure, we illuminate the atoms with one beam to image only the population in the $|F=1\rangle$ state. After a 10-$\mu$s delay, we switch on the same frequency with orthogonal polarization for 90\,$\mu$s and turn on the 3D MOT cooling light. The cooling light depumps all atoms from $|F=2\rangle$ to $|F=1\rangle$ thus allowing us to detect the sum of the two states' populations. The second image forms on the other side of the CCD, due to its deflection at the Wollaston prism. We allow the atoms to disperse and take a second (background) exposure with the same pulse sequence to generate side-by-side absorption images of the $|F=1\rangle$ population and the entire sample. The ratio of the two absorption imaging signals gives $P_{F=1}$. The extinction ratio in each polarization after the Wollaston is $\sim$10,000:1, resulting in negligible crosstalk between the images. With one of the  imaging beams blocked, no signal remains detectable due to the other beam in the final image.


\begin{thebibliography}{1}%
\makeatletter
\providecommand \@ifxundefined [1]{%
 \@ifx{#1\undefined}
}%
\providecommand \@ifnum [1]{%
 \ifnum #1\expandafter \@firstoftwo
 \else \expandafter \@secondoftwo
 \fi
}%
\providecommand \@ifx [1]{%
 \ifx #1\expandafter \@firstoftwo
 \else \expandafter \@secondoftwo
 \fi
}%
\providecommand \natexlab [1]{#1}%
\providecommand \enquote  [1]{``#1''}%
\providecommand \bibnamefont  [1]{#1}%
\providecommand \bibfnamefont [1]{#1}%
\providecommand \citenamefont [1]{#1}%
\providecommand \href@noop [0]{\@secondoftwo}%
\providecommand \href [0]{\begingroup \@sanitize@url \@href}%
\providecommand \@href[1]{\@@startlink{#1}\@@href}%
\providecommand \@@href[1]{\endgroup#1\@@endlink}%
\providecommand \@sanitize@url [0]{\catcode `\\12\catcode `\$12\catcode
  `\&12\catcode `\#12\catcode `\^12\catcode `\_12\catcode `\%12\relax}%
\providecommand \@@startlink[1]{}%
\providecommand \@@endlink[0]{}%
\providecommand \url  [0]{\begingroup\@sanitize@url \@url }%
\providecommand \@url [1]{\endgroup\@href {#1}{\urlprefix }}%
\providecommand \urlprefix  [0]{URL }%
\providecommand \Eprint [0]{\href }%
\providecommand \doibase [0]{http://dx.doi.org/}%
\providecommand \selectlanguage [0]{\@gobble}%
\providecommand \bibinfo  [0]{\@secondoftwo}%
\providecommand \bibfield  [0]{\@secondoftwo}%
\providecommand \translation [1]{[#1]}%
\providecommand \BibitemOpen [0]{}%
\providecommand \bibitemStop [0]{}%
\providecommand \bibitemNoStop [0]{.\EOS\space}%
\providecommand \EOS [0]{\spacefactor3000\relax}%
\providecommand \BibitemShut  [1]{\csname bibitem#1\endcsname}%
\let\auto@bib@innerbib\@empty
%</preamble>
\bibitem [{Note1()}]{Note1}%
  \BibitemOpen
  \bibinfo {note} {We call this term a two-photon detuning loosely. It is the
  difference between the Raman laser frequency difference $\omega _1-\omega _2$
  and the hyperfine splitting $\omega _A-\omega _B$ during free evolution (see
  Fig. \ref {trajectories}(b)). That is, our so-called two-photon detuning does
  not include AC Stark shifts of the internal hyperfine energies induced when
  the Raman light dresses the atoms, but it does include other shifts that
  persist during free evolution such as Zeeman shifts.}\BibitemShut {Stop}%
\end{thebibliography}%


\begin{thebibliography}{99}

%\bibitem{Carnal} O. Carnal and J. Mlynek, %Young's double-slit experiment with atoms: a simple atom interferometer. 
%Phys. Rev. Lett. {\bf 66,} 2689 (1991).

%\bibitem{Pritchard} D. W. Keith, C. R. Ekstrom, Q. A. Turchette, and D. E. Pritchard, %An interferometer for atoms
%Phys. Rev. Lett. {\bf 66,} 2693 (1991). 

\bibitem{CroninReview} A. D. Cronin, J. Schmiedemayer, and D. E. Pritchard, % Optics and interferometry with atoms and molecules
Rev. Mod. Phys. {\bf 81}, 1051 (2009).
    

\bibitem{Peters} A. Peters, K.-Y. Chung, and S. Chu, Nature {\bf 400,} 849 (1999). %; Metrologia {\bf 38,} 25-61 (2001).%6
    
\bibitem{Geiger} R. Geiger,	V. M\'enoret, G. Stern,	N. Zahzam, P. Cheinet, B. Battelier, A. Villing, F. Moron, M. Lours, Y. Bidel, A. Bresson, A. Landragin, and P. Bouyer, %Detecting inertial effects with airborne matter-wave interferometry.
Nat. Commun. {\bf 2,} 474 (2011).
    
\bibitem{McGuirk} J. M. McGuirk, G. T. Foster, J. B. Fixler, M. J. Snadden, and M. A. Kasevich, %Sensitive absolute-gravity gradiometry using atom interferometry,
Phys. Rev. A {\bf 65,} 033608 (2002).

\bibitem{TinoG} G. Rosi, F. Sorrentino, F. Cacciapuoti, M. Prevedelli, Nature {\bf 510,} 518 (2014).

\bibitem{LVGrav} K.-Y. Chung, S.-w. Chiow, S. Herrmann, S. Chu, and H. M\"uller, %Atom interferometry tests of local Lorentz invariance in gravity and electrodynamics
Phys. Rev. D {\bf 80,} 016002 (2009).

\bibitem{Hohensee2012} M. A. Hohensee, B. Estey, P. Hamilton, A. Zeilinger, and H. M\"uller. %Force-free gravitational redshift: a Gravitational Aharonov-Bohm experiment. 
Phys. Rev. Lett. {\bf 108,} 230404 (2012).

\bibitem{EEP1} S. Fray, C.A. Diez, T.W. H\"ansch, and M. Weitz, %Atomic Interferometer with Amplitude Gratings of Light and Its Applications to Atom Based Tests of the Equivalence Principle
Phys. Rev. Lett. {\bf 93,} 240404 (2004).

\bibitem{EEP2} H. M\"uller, A. Peters, and S. Chu, %A precision measurement of the gravitational redshift by the interference of matter waves
Nature {\bf 463,} 926 (2010).

\bibitem{EEP3} A. Bonnin, N. Zahzam, Y. Bidel, and A. Bresson, % Simultaneous dual-species matter-wave accelerometer
Phys. Rev. A {\bf 88,} 043615 (2013).

\bibitem{EEP4} D. Schlippert, J. Hartwig, H. Albers, L. L. Richardson, C. Schubert, A. Roura, W. P. Schleich, W. Ertmer, and E.  M. Rasel, %Quantum Test of the Universality of Free Fall
Phys. Rev. Lett. {\bf 112,} 203002 (2014).

\bibitem{EEP5}M. G. Tarallo, T. Mazzoni, N. Poli, D. V. Sutyrin, X. Zhang, and G. M. Tino, %Test of Einstein Equivalence Principle for 0-Spin and Half-Integer-Spin Atoms: Search for Spin-Gravity Coupling Effects
Phys. Rev. Lett. {\bf 113,} 023005 (2014).

\bibitem{EEP6}L. Zhou, S. Long, B. Tang, X. Chen, F. Gao, W. Peng, W. Duan, J. Zhong, Z. Xiong, J. Wang, Y. Zhang, and M. Zhan, %Test of Equivalence Principle at 10^−8 Level by a Dual-Species Double-Diffraction Raman Atom Interferometer
Phys. Rev. Lett. {\bf 115,} 013004 (2015).

\bibitem{EEP7}X.-C. Duan, X.-B. Deng, M.-K. Zhou, K. Zhang, W.-J. Xu, F. Xiong, Y.-Y. Xu, C.-G. Shao, J. Luo, and Z.-K. Hu, Phys. Rev. Lett. {\bf 117,} 023001 (2016).

\bibitem{Chameleons} P. Hamilton, M. Jaffe, P. Haslinger, Q. Simmons, H. M\"uller, and J. Khoury, Science {\bf 349,} 849 (2015).%14
%B. Elder, {\em et al.,} arXiv:1603.06587

\bibitem{CCC} S.-Y. Lan, P.-C. Kuan, B. Estey, D. English, J. Brown, M. Hohensee, and H. M\"uller, Science {\bf 339,} 554 (2013).

\bibitem{Bouchendira2013} R. Bouchendira, P. Clad\'e, S. Guellati-Kh\'elifa, F. Nez, and F. Biraben, Ann. Phys. (Berlin) {\bf 525,} 484 (2013).

%\bibitem{Weiss1993} D. S. Weiss, B. C. Young, and S. Chu, Phys. Rev. Lett. {\bf 70,} 2706 (1993).
 
\bibitem{Weiss1994} D. S. Weiss, B. C. Young, and S. Chu, %Precision measurement of h/mcs based on photon recoil using laser-cooled atoms and atomic interferometry,
    Appl. Phys. B. {\bf 59,} 217 (1994).  
 
\bibitem{Wicht} A. Wicht, J. M. Hensley, E. Sarajlic, and S. Chu, Phys. Scr. {\bf T102,} 82 (2002).

\bibitem{Bouchedira2011} R. Bouchendira, P. Clade, S. Guellati-Khelifa, F. Nez, and F. Biraben, Phys. Rev. Lett. {\bf 106,} 080801 (2011).

\bibitem{Jamison} A. O. Jamison, B. Plotkin-Swing, and S. Gupta, %Advance in precision contrast interferometry with Yb Bose-Einstein condensates. 
Phys. Rev. A {\bf 90,} 063606 (2014).

\bibitem{Estey2015} B. Estey, C. Yu, H. M\"uller, P.-C. Kuan, and S.-Y. Lan, Phys. Rev. Lett. {\bf 115,} 083002 (2015). 

\bibitem{Gabrielse} G. Gabrielse, D. Hanneke, T. Kinoshita, M. Nio, and B. Odom,
Phys. Rev. Lett. {\bf 97,} 030802 (2006). %; Erratum Phys. Rev. Lett. 99, 039902 (2007)New Determination of the Fine Structure Constant from the Electron g Value and QED

\bibitem{PenningTrapSME} Y. Ding and V. A. Kostelecky, Phys. Rev. D {\bf 94,} 056008 (2016).

\bibitem{Miffre} C. Champenois, M. B\"uchner, and J. Vigu\'e, % Fringe contrast in three grating Mach-Zehnder atomic interferometers. 
 Eur. Phys. J. D {\bf 5,} 363-374 (1999).
 
\bibitem{Jac} M. Jacquey, M. B\"uchner, G. Tr\'enec, and J. Vigu\'e, % First measurements of the index of refraction of gases for lithium atomic waves. 
 Phys. Rev. Lett. {\bf 98,} 240405 (2007).

\bibitem{SCI} S.-w. Chiow, S. Herrmann, S. Chu, and H. M\"uller, %Noise-Immune Conjugate Large-Area Atom Interferometers,
Phys. Rev. Lett. {\bf 103,} 050402 (2009).

\bibitem{Noisered} S.-w. Chiow, J. Williams, and N. Yu, Phys. Rev. A {\bf 93,} 013602 (2016).

\bibitem{Bayanal} J. K. Stockton, X. Wu, and M. A. Kasevich,
Phys. Rev. A {\bf 76,} 033613 (2007).

\bibitem{KasevichChu} M. Kasevich and S. Chu, Phys. Rev. Lett. {\bf 67,} 181 (1991).

\bibitem{BiedermannWarm} G. W. Biedermann, H. J. McGuinness, A. V. Rakholia, Y.-Y. Jau, D. R. Wheeler, J. D. Sterk, and G. R. Burns, %. Atom Interferometry in a Warm Vapor.
arXiv:1610.02451.

\bibitem{AllOptical} P. M. Duarte, R. A. Hart, J. M. Hitchcock, T. A. Corcovilos, T. -L. Yang, A. Reed, and R. G. Hulet, Phys. Rev. A {\bf 84,} 061406 (2011). 

\bibitem{vsel} M. Kasevich, D. S. Weiss, E. Riis, K. Moler, S. Kasapi, and S. Chu, Phys. Rev. Lett. {\bf 66,} 2297 (1991).

\bibitem{nuclear} M. A. Hohensee, H. M\"uller, and R. B. Wiringa, %Equivalence Principle and Bound Kinetic Energy. http://prl.aps.org/edannounce/PhysRevLett.98.010001
Phys. Rev. Lett. {\bf 111,} 151102 (2013).

\bibitem{eCooling} A. P. Povilus, N. D. DeTal, L. T. Evans, N. Evetts, J. Fajans, W. N. Hardy, E. D. Hunter, I. Martens, F. Robicheaux, S. Shanman, C. So, X. Wang, and J. S. Wurtele, Phys. Rev. Lett. {\bf 117,} 175001 (2016). 

\bibitem{eLorentz} K.-P. Marzlin and T. Lee, Phys. Rev. A {\bf 89,} 062103 (2014).

%\bibitem{MolecularFountain} C. Cheng, A. P. P. van der Poel, P. Jansen, M. Quintero-P\'erez, T. E. Wall, W. Ubachs, and H. L. Bethlem, Phys. Rev. Lett. {\bf 117,} 253201 (2016).

%\bibitem{ALPHA} The ALPHA Collaboration, Nature Phys. {\bf 7,} 558-564 (2011). 

%\bibitem{Treutlein2001} P. Treutlein, K.-Y. Chung, and S. Chu, % High-brightness atom source for atomic fountains. 
% Phys. Rev. A {\bf 63,} 051401(R) (2001).

 %; Delhuille, R; Champenois, C; Buchner, M; et al., High-contrast Mach-Zehnder lithium-atom interferometer in the Bragg regime. Appl. Phys. B {\bf 74,}, 489-493 (2002); Miffre, A; Jacquey, M; Buchner, M; et al., Lithium atom interferometer using laser diffraction: description and experiments. Eur. Phys. J. D {\bf 33,}, 99-112 (2005); Miffre, A; Jacquey, M; Buchner, M; et al., Measurement of the electric polarizability of lithium by atom interferometry, Phys. Rev. A {\bf 73,} 011603 (2006) %;  Miffre, A; Jacquey, M; Buchner, M; et al., Atom interferometry measurement of the electric polarizability of lithium. Eur. Phy. J. D {\bf 38,} 353-365 (2006); 

%; Jacquey, M.; Miffre, A.; Buechner, M.; et al., Test of the isotopic and velocity selectivity of a lithium atom interferometer by magnetic dephasing. EPL {\bf 77,} 20007 (2007); Jacquey, M.; Miffre, A.; Buechner, M.; et al., Phase noise due to vibrations in Mach-Zehnder atom interferometers, EUROPHYSICS LETTERS {\bf 75,} 688-694 (2006)

\bibitem{HoganLPAI} J. M. Hogan, D. M. S. Johnson, and M. A. Kasevich, %. Light-pulse atom interferometry.
arXiv:0806.3261.

\bibitem{LIcooling}  P. Hamilton, G. Kim, T. Joshi, B. Mukherjee, D. Tiarks, and H. M\"uller, %Sisyphus Cooling of Lithium. 
Phys. Rev. A {\bf 89,} 023409 (2014). % arxiv:1308.1935.

\bibitem{LiOP} J. Gillot, A. Gauguet, M. B\"uchner, and J. Vigu\'e, % Optical pumping of a lithium atomic beam for atom interferometry
Eur. Phys. J. D {\bf 67,} 263 (2013).

\bibitem{GrayMolasses} A. T. Grier, I. Ferrier-Barbut, B. S. Rem, M. Delehaye, L. Khaykovich, F. Chevy, and C. Salomon, Phys. Rev. A {\bf 87,} 063411 (2013).

\bibitem{Multiport} R. H. Parker, C. Yu, B. Estey, W. Zhong, E. Huang, and H. M\"uller, % Controlling the multiport nature of Bragg diffraction in atom interferometry
Phys. Rev. A {\bf 94,} 053618 (2016).

\bibitem{Paul} P. Hamilton, M. Jaffe, P. Haslinger, Q. Simmons, H. Müller, and J. Khoury, Science {\bf 349,} 849-851 (2015).

\bibitem{eRBI} K.-P. Marzlin, Phys. Rev. A {\bf 88,} 043621 (2013).

\bibitem{McGregorBichromatic} S. McGregor, W. C.-W. Huang, B. A. Shadwick, and H. Batelaan, Phys. Rev. A {\bf 92,} 023834 (2015).

\bibitem{Biedermann14} A. V. Rakholia, H. J. McGuinness, and G. W. Biedermann, %Dual-Axis High-Data-Rate Atom Interferometer via Cold Ensemble Exchange, 
Phys. Rev. Applied {\bf 2,} 054012 (2014).

% ==== not cite in text ==== %
% \bibitem{Asenbaum} P. Asenbaum, C. Overstreet, T. Kovachy, D. D. Brown, J. M. Hogan, M. A. Kasevich, arXiv:1610.03832v2.

% \bibitem{Biedermann} H. J. McGuinness, A. V. Rakholia, and G. W. Biedermann, %High data-rate atom interferometer for measuring acceleration
% Appl. Phys. Lett. {\bf 100,} (2012).

% 

% \bibitem{AntiHInt} P. Hamilton, A. Zhmoginov, F. Robicheaux, J. Fajans, J. S. Wurtele, and H. M\"uller, %'Antimatter interferometry for gravity measurements. 
% Phys. Rev. Lett. {\bf 112,} 121102 (2014). 
    
% \bibitem{Gillot} J. Gillot, A. Gauguet, M. Buechner {\em et al.}, %Optical pumping of a lithium atomic beam for atom interferometry By:  
% Eur. Phys. J. {\bf 67,} 263 (2013).

%\bibitem{Lepoutre} Test of the He-McKellar-Wilkens topological phase by atom interferometry. II. The experiment and its results By: Lepoutre, S.; Gillot, J.; Gauguet, A.; et al. PHYSICAL REVIEW A  Volume: 88   Issue: 4     Article Number: 043628   Published: OCT 21 2013

% \bibitem{Hamilton2015} P. Hamilton, M. Jaffe, J. M. Brown, L. Maisenbacher, B. Estey, and H. M\"uller, %Atom interferometry in an optical cavity. 
% Phys. Rev. Lett. {\bf 114,} 100405 (2015).

% \bibitem{Barrett} B. Barrett, A. Carew, S. Beattie, and A. Kumarakrishnan, %Measuring the atomic recoil frequency using a modified grating-echo atom interferometer
% Phys. Rev. A {\bf 87,} 033626 (2013) %Rb interferometer reaches 37 ppb in h/m in in 14 h, but 6 ppm systematics

% \bibitem{Gupta} S. Gupta, K. Dieckmann, Z. Hadzibabic, and D.E. Pritchard, %Contrast Interferometry using Bose-Einstein Condensates to Measure h/m andα, 
% Phys. Rev. Lett. {\bf 89,} 140401 (2002).



\end{thebibliography}
\end{document}